\documentstyle{livrev}
\newcommand{\bea}{\begin{eqnarray}}
\newcommand{\eea}{\end{eqnarray}}

\def\R{{\rm I\!R}}

\def\C{{\mathchoice
{\setbox0=\hbox{$\displaystyle\rm C$}\hbox{\hbox to0pt
{\kern0.4\wd0\vrule height0.9\ht0\hss}\box0}}
{\setbox0=\hbox{$\textstyle\rm C$}\hbox{\hbox to0pt
{\kern0.4\wd0\vrule height0.9\ht0\hss}\box0}}
{\setbox0=\hbox{$\scriptstyle\rm C$}\hbox{\hbox to0pt
{\kern0.4\wd0\vrule height0.9\ht0\hss}\box0}}
{\setbox0=\hbox{$\scriptscriptstyle\rm C$}\hbox{\hbox to0pt
{\kern0.4\wd0\vrule height0.9\ht0\hss}\box0}}}}

\font\fivesans=cmss10 at 4.61pt
\font\sevensans=cmss10 at 6.81pt
\font\tensans=cmss10
\newfam\sansfam
\textfont\sansfam=\tensans\scriptfont\sansfam=\sevensans\scriptscriptfont
\sansfam=\fivesans
\def\sans{\fam\sansfam\tensans}
\def\Z{{\mathchoice
{\hbox{$\sans\textstyle Z\kern-0.4em Z$}}
{\hbox{$\sans\textstyle Z\kern-0.4em Z$}}
{\hbox{$\sans\scriptstyle Z\kern-0.3em Z$}}
{\hbox{$\sans\scriptscriptstyle Z\kern-0.2em Z$}}}}

\begin{document}


\title{\hfill {\large AEI-063}\\
\hfill {\large May 1998}\\ \hfill {} \\ \hfill {} \\
Discrete approaches to quantum gravity in four dimensions}

\author{Renate Loll\\
Max-Planck-Institut f\"ur Gravitationsphysik\\
Schlaatzweg 1\\
D-14473 Potsdam\\
e-mail: loll@aei-potsdam.mpg.de}

\date{}
\maketitle

\thanks{\center This article is an invited contribution to Living Reviews,\\
\hspace{3.5cm} and will appear at http://www.livingreviews.org/.}

\vfill

\begin{center}
{\bf Abstract}
\end{center}
The construction of a consistent theory of quantum gravity
is a problem in theoretical physics that has so far
defied all attempts at resolution. One ansatz 
to try to obtain a non-trivial quantum theory proceeds via a
discretization of space-time and the Einstein action. 
I review here three major areas of research: 
gauge-theoretic approaches,
both in a path-integral and a Hamiltonian formulation, quantum
Regge calculus, and the method of dynamical triangulations, 
confining attention to work that is strictly four-dimensional,
strictly discrete, and strictly quantum in nature. 

\vfill

\newpage

\section{Introduction}

With the absence of a satisfactory quantum theory of gravity, 
a major puzzle in theoretical physics remains still unsolved. 
This article contains an overview and a comprehensive
bibliography 
of past efforts to define a consistent theory of
quantum gravity in four dimensions via an intermediate
discretization. 
I will discuss only models with some concrete implementation
of the dynamics of Einstein's theory, Lagrangian or
Hamiltonian. 

One way of tackling the quantization problem 
non-perturbatively is to use discrete methods, in analogy 
with quantum field theories on a flat background. 
They of course have to be
modified in the case of gravity, where the metric of space-time 
becomes itself a dynamical variable. The diffeomorphism
invariance of the classical continuum theory is no obstacle
in principle to the introduction of a discretization, as has 
been demonstrated by the success of such methods in
describing two-dimensional Euclidean gravity in a path-integral
approach. 

The approaches I will describe below have yielded a variety of
interesting results on the discrete simulation methods themselves,
as well as on the geometric properties of typical configurations,
the role of gauge invariance, the phase structure, the
inclusion of matter, to name just a few.
They have not so far been successful 
in providing convincing evidence for the existence of
a non-trivial four-dimensional quantum theory of gravity (neither,
of course, have other methods). There are considerable 
technical difficulties in performing 
analytical computations and setting up sufficiently big
numerical simulations.
It is a theoretical challenge to come up with more
realistic models that may lead to a truly interacting 
gravitational theory. 

I have divided the discrete approaches into three categories,
which I will discuss in the order of their chronological
appearance:
the gauge-theoretic formulations using connection variables, 
based on first-order descriptions of Einstein gravity, and
two metric formulations, the quantum Regge calculus program and 
the more recent method of dynamical triangulations, which use
simplicial instead of hypercubic lattice discretizations.
They mainly deal with pure gravity,
with the possible inclusion of a cosmological constant
term and higher-order derivative terms, and in some instances,
matter-coupling.

This review covers about 200 papers, the earliest
of which appeared in 1979 (not counting the 1964 article by 
Leutwyler \cite{Leutwyler1964},
who used a lattice approximation to perform a gravitational
sum over histories \`a la Feynman). Much of the motivation for the
discrete investigations comes of course from other sources,
which for the most part I do not cite explicitly. 
The reader is referred to the individual articles 
for a more complete bibliography of the background material.

\section{Gauge-theoretic discretizations of gravity}

\subsection{Lagrangian treatment. Introduction}
This area of research was inspired by
the success of non-perturbative lattice methods in
treating non-abelian gauge theories \cite{Rebbi1983}.
To apply some of their
techniques, gravity has to be brought into a gauge-theoretic,
first-order form, with the pure-gravity Lagrangian
\begin{equation}\label{Z.1}
S[A,e]= \int_{M} e\wedge e\wedge R[A],
\end{equation}
\noindent where the $so(3,1)$-valued spin-connection $A_\mu^{IJ}$ 
(with curvature $R$) and the vierbein 
$e_\mu^I$ are considered as independent variables. The important
feature of (\ref{Z.1}) is that $A_\mu^{IJ}$
is a gauge potential, and that the 
action -- in addition to its diffeomorphism-invariance -- 
is invariant under local frame rotations. 
Variation with respect to $e_\mu^I$ leads to the
metricity condition,
\begin{equation}\label{Z.2}
{\cal D}_{[\mu} e_{\nu]}^I\equiv \partial_{[\mu} e_{\nu]}^I+
A_{[\mu\; J}^{\; I}\, e_{\nu]}^J=0,
\end{equation} 
\noindent which can be solved to yield the unique torsion-free
spin connection $A=A[e]$ compatible with the $e_{\mu I}$.
There are some obvious differences with usual gauge theories:
i) the action (\ref{Z.1}) is linear instead of quadratic in the curvature 
two-form $R_{\mu\nu}^{IJ}$ of $A$, and 
ii) it contains additional fields $e_\mu^I$. 
Substituting the solution to (\ref{Z.2}) into the action, one obtains 
$S[e]=\int_{M} d^{4}x\, (\det e)\, R$,
where $R$ denotes the four-dimensional curvature scalar. 
This expression coincides with the usual Einstein action
$\int d^{4}x\, \sqrt{\det g}\, R$ only for $\det e >0$.

Most of the lattice gauge formulations I will discuss below 
share some common features.
The lattice geometry is hypercubic, defining
a natural global coordinate system
for labelling the lattice sites and edges. 
The gauge group $G$ is $SO(3,1)$ or
its ``Euclideanized" form $SO(4)$, or a larger group
containing it as a subgroup or via a contraction limit. 
Local curvature terms are represented by (the traces of) $G$-valued Wilson 
holonomies $U_{\sqcap \mskip-12mu \sqcup}$ around lattice
plaquettes. The vierbeins are either considered as additional
fields or identified with part of the connection variables.
The symmetry group of the lattice Lagrangian is 
a subgroup of the gauge group $G$, and does not
contain any translation generators that appear when $G$ is the
Poincar\'e group.

When discretizing conformal gravity (where $G=SO(5,1)$) or higher-derivative
gravity in first-order form, the metricity condition on the 
connection has to be imposed by hand. This leads to technical complications 
in the evaluation of the functional integral. 

The diffeomorphism invariance of the continuum theory
is broken on the lattice; only the local gauge
invariances can be preserved exactly. The reparametri\-za\-tion invariance
re-emerges only at the linearized level, i.e. when considering
small perturbations about flat space. 

\subsection{Smolin's lattice model}
The first gauge-theoretic model for lattice gravity 
is due to Smolin \cite{Smolin1979}, based on the continuum formulation of
MacDowell and Mansouri \cite{MacDowellMansouri1977}, 
with de Sitter gauge group $O(3,2)$ or 
$O(4,1)$, and a Lagrangian of $R^2$-type,
\begin{equation}\label{Z.3}
S= \int d^4x\; \epsilon^{\mu\nu\rho\sigma} \epsilon_{IJKL}\,
\tilde R_{\mu\nu}^{IJ}\, \tilde R_{\rho\sigma}^{KL},
\end{equation}
\noindent where the components of $\tilde R_{\mu\nu}$ are related
to those of the usual curvature tensor by
\begin{equation}\label{Z.4}
\tilde R_{\mu\nu}^{IJ}=
R_{\mu\nu}^{IJ}\pm\frac{1}{l^{2}}
(e_{\mu}^{I}e_{\nu}^{J}-e_{\mu}^{J}e_{\nu}^{I}).
\end{equation}
\noindent Although the underlying gauge potentials $A_\mu$ are
$o(3,2)$- or $o(4,1)$-valued, the action is only invariant under
the 6-dimensional subgroup of Lorentz transformations. The theory
contains a dimensionful parameter $l$.
The gauge potentials associated with the internal 5-direction are 
identified with the frame fields $e_\mu^I$, and the 
action can be decomposed
into the usual Einstein-term (\ref{Z.1}) plus a cosmological constant term
with $\lambda\sim\frac{1}{l^{2}}$ and a topological $R\wedge R$-term. 
Smolin analyzed its lattice discretization, and found
both a weak- and a strong-coupling phase, with respect to the dimensionless
coupling constant $g\sim\frac{\sqrt{G}}{l}$. He performed a 
weak-coupling expansion about
flat space and rederived the usual propagator. In the
strong-coupling regime he found massive excitations and a confining 
property for spinors.

\subsection{Numerical implementation}
Caracciolo and Pelissetto 
\cite{CaraccioloPelissetto1987a,CaraccioloPelissetto1987b,
CaraccioloPelissetto1988a,CaraccioloPelissetto1988b, 
CaraccioloPelissetto1989} performed a numerical 
investigation of the phase structure of Smolin's model. Using 
the compact group $SO(5)$ and its associated
Haar measure, their findings confirmed the two-phase structure:
a strong-coupling phase with a confining property and presence of
exponential clustering, and a weak-coupling phase dominated by
a class of topological configurations, with vanishing vierbein.
However, their Monte Carlo data (on $4^{4}$ and $8^{4}$-lattices with
periodic boundary conditions) indicated strongly that the transition
was first-order, even if the measure was generalized 
by a factor of $| \det e |^{p}$, $p\in [0,150]$ 
\cite{CaraccioloPelissetto1989}.

\subsection{Other gauge formulations}
Das et al \cite{Dasetal1979} lattice-discretized an 
$Sp(4)$-invariant Lagrangian due to West \cite{West1978}, 
leading to a functional form
$S[U]=\sum_{n}\sqrt{{\rm Tr}\;((U_{n}-\frac{1}{4}{\rm 
Tr}\;U_{n})^{2})}$, where $U_{n}$ denotes a sum of double-plaquette 
holonomies based at the vertex $n$.
Details of the functional integration
were not spelled out. 
The square-root form of the Lagrangian does not
make it very amenable to numerical investigations (see also the 
comments in \cite{Kaku1987}).

Mannion and Taylor \cite{MannionTaylor1981} suggested a discretization of the
$(\int e\wedge e\wedge R)$-Lagrangian without cosmological term,
with the vierbeins $e$ treated as extra variables, and the compac\-tified
Lorentz group $SO(4)$. 

Kaku \cite{Kaku1983} has proposed a lattice 
version of conformal gravity (see also \cite{Kaku1981,Kaku1987}),
based on the group $O(4,2)$, the main hope being that unitarity
could be demonstrated non-perturbatively. 
As usual, a metricity constraint on the 
connection has to be imposed by hand. 

A lattice formulation of higher-derivative gravity, containing 
fourth-order $R^{2}$-terms was given by Tomboulis \cite{Tomboulis1984}. 
The continuum
theory is renormalizable and asymptotically free, but has problems
with unitarity. The motivation for this work was again the hope of realizing
unitarity in a lattice setting. 
The square-root form of the
Lagrangian is similar to that of Das et al. 
It is
$O(4)$-invariant and supposedly satisfies reflection positivity. 
Again, the form of the Lagrangian and the measure (containing
a $\delta$-function of the no-torsion constraint) is rather 
complicated and has not been used for a further non-perturbative
analysis.

Kondo \cite{Kondo1984} employed the same framework as Mannion and Taylor, but 
introduced an explicit symmetrization of the Lagrangian. 
He claimed that the cluster expansion goes through just as
in lattice Yang-Mills theory, leading to a positive mass gap.

\subsection{Proving reflection positivity}
Various properties of gauge gravity models were 
analyzed by Menotti and Pelissetto in a series of papers in the 
mid-eighties. 
They first studied a discrete $O(4)$-version 
of the Lagrangian (\ref{Z.1}) \cite{MenottiPelissetto1986}, 
including the $O(4)$-Haar measure and
a general local $O(4)$-invariant measure for the vierbein fields, 
and showed that reflection-positivity holds only for a restricted
class of functions. Furthermore, expanding 
about flat space, and after appropriate gauge-fixing, they discovered 
a doubling phenomenon similar to that found for chiral fermions in
lattice gauge theory, a behaviour that also
persists for different gauge-fixings. 
One finds the same mode doubling
also for a flat-background expansion of conformal gravity 
\cite{MenottiPelissetto1987b}.
In the same paper, they gave a unified treatment of
Poincar\'e, de Sitter and conformal gravity, and showed 
that reflection positivity
for $O(4)$-gravity (as well as for the two other gauge groups) 
holds exactly and for general functions only provided a
signature factor sign$(\det e)$ is included in the Lagrangian. 

To ensure the convergence of the functional integration, one has to
introduce a damping factor for the vierbeins in the measure, both for
Poincar\'e and conformal gravity \cite{MenottiPelissetto1987b}. 
An extension of the results of \cite{MenottiPelissetto1987b} 
to supergravity
with the super-Poincar\'e group is also possible \cite{Caraccioloetal1988}. 
One can prove
reflection positivity and finds a matching gravitino doubling in the
perturbative expansion.

\subsection{The measure}
One can derive non-trivial conditions on the lattice measure by imposing the
Slav\-nov-Taylor identities in a perturbative lattice calculation. 
Starting from the general form of the measure for Poincar\'e gravity,
\begin{equation}\label{Z.5}
\prod_{n\mu\nu\rho\sigma} \det ( e_{\mu}(n),e_{\nu}(n),e_{\rho}(n),
e_{\sigma}(n) )^{N/16}\prod_{n\mu}f(e_{\mu}^{2}(n))
\prod_{n,\mu >0}de_{\mu}(n) \prod_{n,\mu >0} dU_{L}(n,n+\mu),
\end{equation}
\noindent Menotti and Pelissetto \cite{MenottiPelissetto1987a}
performed a one-loop calculation
and found that the Slavnov-Taylor identity can be satisfied 
for a particular choice of $N$, $f_{1},f_{2}$ (momentum expansion 
coefficients of $f$), and of the cosmological constant
$\lambda$. The solution still depends on a real parameter $\xi$
(related to a residual non-invariance under rotations). 
This makes it difficult to draw any immediate conclusions
on the structure of the full, non-perturbative measure.

\subsection{Assorted topics}
Caselle et al \cite{Caselleetal1987a} 
proposed a lattice action that is genuinely 
Poincar\'e-invariant, at the price of introducing additional
lattice ``coordinate variables''.
They also suggested a compact
$O(5)$-formulation which reduces to the Poincar\'e
form in the limit as the length of some preferred $O(5)$-vector 
is taken to infinity, as well as a super-version involving the
graded Poincar\'e group. 
The same authors in \cite{Caselleetal1987b} put forward 
an argument 
for why doubling should appear in general gravity plus matter systems.

Reisenberger \cite{Reisenberger1997} 
has recently suggested a gauge-theoretic
path integral based on the Plebanski action for Euclidean
gravity. He discretizes the theory on a simplicial or hypercubic
lattice with group- and algebra-valued fields. 
A metricity constraint needs to be imposed on the
basic spin-1 fields, which it turns out is difficult to treat exactly.

\subsection{Summary}
The gauge-theoretic Lagrangian lattice approaches are afflicted
by a number of technical difficulties. 
Reflection positivity can be shown for some of the models, but
generally requires the inclusion of a factor sign$(\det e)$ in the
Lagrangian. 
Obtaining qualitative non-perturbative information
about the phase structure requires a non-trivial measure input. The
complicated functional form of the Lagrangian and the metri\-city condition
that has to be imposed via a Lagrange multiplier and the corresponding
functional integration do not make conformal and higher-derivative
theories attractive candidates for numerical simulations. 
The compact version of Smolin's de Sitter gravity is
still the simplest model, but its numerical investigation 
did not yield interesting results.

\subsection{Hamiltonian treatment. Introduction} 
Relatively little work has been done on discretized Hamiltonian
fomulations of gravity. This can in part be understood from the
fact that the numerical methods available
for lattice gauge field theories rely mostly on the Euclidean path-integral
description. 
Unfortunately the relation between the Lagrangian and
Hamiltonian quantizations for generally covariant theories without a
fixed background is far from clear. 

In the usual metric formulation, the complicated non-polynomial
form of the Hamiltonian constraint has been a long-standing problem.
In this framework, neither the functional form of the quantum 
representations nor the nature of the quantization problems 
suggest that a discrete approach might yield any advantages. 
This situation has improved with the introduction of new Hamiltonian
gauge-theoretic variables by Ashtekar \cite{Ashtekar1986,Ashtekar1987}.

By a Hamiltonian lattice approach one usually means a formulation in 
which the time variable is left continuous, and only the spatial
3-slices are discretized. 
In continuum gravity, the 3+1 decomposition 
leads to the (non-Lie) Dirac algebra of
the three-dimensional diffeomorphism generators and the
Hamiltonian constraint, 
associated with the deformation of three-surfaces imbedded 
in four-space. 
One usually requires 
this algebra to be realized in the quantum theory, without anomalous terms, 
for a set of
self-adjoint quantum constraint operators, for some factor-ordering.

Since a discretization of space-time breaks the diffeomorphism invariance, 
there is no reason to expect the Dirac algebra to be preserved
in any discrete approach, even classically. 
This raises the question of whether and in what form part of the 
diffeomorphism symmetry can still be realized at the discrete level.
Using gauge-theoretic variables, one can maintain the
exact local gauge invariance with respect to the internal degrees
of freedom, but there is no analogous
procedure for treating the coordinate invariance.

\subsection{Hamiltonian lattice gravity}
There is a gauge-theoretic Hamiltonian version of gravity defined on a 
cubic lattice, which in many aspects resembles the Lagrangian gauge
formulations described earlier. It also is virtually the only
discrete Hamiltonian formulation in which some progress has
been achieved in the quantization (see also \cite{Loll1997d} for a recent
review). 

Renteln and Smolin \cite{RentelnSmolin1989} were the first to set up a 
continuous-time lattice discretization along the lines of 
Hamiltonian lattice gauge theory. Their basic configuration
variables are the link holonomies $U(l)$ of the spatial Ashtekar 
connection along the edges. 
The lattice analogues of the canonically
conjugate pairs $(A_{a}^{i}(x),E^{a}_{i}(x))$ are the link variables
$(U(l)_{A}{}^{B},p_{i}(l))$, with Poisson brackets
\begin{eqnarray}\label{Z.6}
&&\{ U_A{}^B(n,\hat a), U_C{}^D(m,\hat b)\} =0,\\
&&\{  p_i(n,\hat a), U_A{}^C(m,\hat b)\} =
-\frac{1}{2}\,\delta_{nm}\delta_{\hat a\hat b}\, \tau_{iA}{}^B U_B{}^C
(n,\hat a),\\
&&\{ p_i(n,\hat a), p_j(m,\hat b) \} =
 \delta_{nm}\delta_{\hat a\hat b}\, \epsilon_{ijk}\,  p_k (n,\hat a),
\end{eqnarray}
\noindent with the $SU(2)$-generators satisfying 
$[\tau_i,\tau_j]=2\,\epsilon_{ijk}\tau_k$. Lattice
links $l=(n,\hat a)$ are labelled by a vertex $n$ and a 
lattice direction $\hat a$.
These relations go over to the usual continuum brackets in the 
limit as the lattice spacing $a$ is taken to zero. 
In this scheme, they wrote down
discrete analogues of the seven polynomial first-class constraints, 
and also attempted to interpret the action of 
the discretized diffeomorphism and Hamiltonian constraints in terms of 
their geometric action on lattice Wilson loop states. 

\subsection{The measure}
There is a natural measure for the quantum
theory, given by the product over all lattice edges
of the Haar measures $dg$.
However, since the Ashtekar connections $A$ are complex-valued, the
gauge group is the non-compact group $SO(3,\C)=SL(2,\C)$, and the 
gauge-invariant Wilson loop functions are not square-integrable. 
For the alternative
formulation in terms of {\it real} SU(2)-variables (see
below), these problems are not present.
An alternative heat kernel measure $d\nu$ for holomorphic 
$SL(2,\C)$ holonomies on the lattice was used in
\cite{Loll1995a,Ezawa1996}.

\subsection{The constraint algebra}
This line of research was continued by Renteln \cite{Renteln1990}, 
who proved that
for a particular factor-ordering (all momenta to the left), the 
subalgebra of the discretized diffeomorphism constraints (smeared by lapse 
functions $N$),
\begin{equation}\label{Z.7}
\sum_{n} {\cal V}^{\rm latt}[N,n):=
\sum_{n} N^{\rm latt}(n,\hat a) \,{\rm 
Tr}\,(U_{\sqcap \mskip-11.5mu \sqcup_{ab}}\tau_{i})p_{i}(n,\hat b),
\end{equation} 
\noindent closes in the
limit as the lattice spacing is taken to zero. This calculation was
later extended to a 
variety of different symmetrizations for the lattice operator and
to an arbitrary factor-ordering of the form
$\alpha\, {\rm Tr}\, (\hat U\tau)\hat p +(1-\alpha ) \hat p\, 
{\rm Tr}\, (\hat U\tau)$, with $0\leq\alpha\leq 1$ \cite{Loll1998}. 
Again, one does not find any quantum anomalies. 
It would be highly desirable to extend this result to commutators
involving also the discretized Hamiltonian constraint
and to find the explicit functional form of the anomalies, if there 
were any.

\subsection{Solutions to the Wheeler-DeWitt equation}
Part of solving the canonical quantum theory is to determine
the states annihilated by the 
Hamiltonian constraint $\hat H$. It was
shown by Loll \cite{Loll1995a} that solutions exist 
in the Renteln-Smolin formulation, where
$\hat H(n)=\sum_{a,b}\epsilon^{ijk} {\rm Tr}\, 
(\hat U(n,{\sqcap \mskip-12mu \sqcup}_{\hat a \hat b})\tau_{k})
\hat p_{i}(n,a)\hat p_{j}(n,b)$. 
They are given by multiple, non-intersecting Polyakov loops
(the lattice is assumed to have compact topology $T^{3}$).
Such solutions are trivial in the sense that
they correspond to quantum states ``without volume''.
The difficulties one encounters when trying
to find other solutions is illustrated by the explicit
calculations for the $1\times 1\times 1$-lattice in \cite{Loll1995a}. 

The search for solutions was continued by Ezawa 
\cite{Ezawa1996} (see also
\cite{Ezawa1997} for an extensive review), who used a
symmetrized form of the Hamiltonian.
His solutions depend on multiple plaquette loops
$(U_{\sqcap \mskip-12mu \sqcup})^k$,
where a single lattice plaquette ${\sqcap \mskip-12mu \sqcup}$
is traversed by the loop
$k$ times. The solutions are less trivial than those formed
from Polyakov loops, since they involve kinks, but they are still
annihilated by the volume operator.

A somewhat different strategy was followed by Fort et al
\cite{Fortetal1996}, who constructed a Hamiltonian
lattice regularization for the calculation of certain knot invariants.
They defined lattice constraint operators
in terms of their geometric action on lattice Wilson loop
states, and reproduced some of the formal
continuum solutions to the polynomial Hamiltonian constraint
of complex Ashtekar gravity on simple loop geometries.

\subsection{The role of diffeomorphisms}
As in other discrete formulations,
the spatial diffeomorphism group cannot be realized exactly
on the lattice, and
the only obvious symmetries of a cubic
lattice are discrete rotations and overall 
translations. 
The commutator computation 
described above indicates that one may be able recover the
diffeomorphism invariance in a suitable continuum limit.
Corichi and Zapata \cite{CorichiZapata1997} have suggested the 
presence of a residual diffeomorphism symmetry in the lattice theory,
under which, for example, all non-intersecting Wilson loop 
lattice states would be identified.

One can try to interpret
the lattice theory as a manifestly 
diffeomorphism-invariant construction, with the lattice
representing an entire diffeomorphism equivalence class
of lattices embedded in the continuum \cite{Loll1995a}. 
In order to make this interpretation consistent, one should
modify the functional form of either the Hamiltonian or the measure,
in such a way that the commutator of two lattice Hamiltonians 
vanishes, as $a\rightarrow 0$.

\subsection{The volume operator}
An important quantity in Hamiltonian lattice
quantum gravity is the volume operator,
the quantum analogue of the classical volume function
$\int d^3x\;\sqrt{\det g}$.
The continuum dreibein determinant $\det E(x)$ (with $|\det E(x)|=\det 
g$), has a natural lattice analogue, given by
\begin{equation}\label{Z.8}
\det p(n):=\frac{1}{3!}\, \epsilon_{abc}\,\epsilon^{ijk} p_i(n,\hat a) 
p_j(n,\hat b) p_k(n,\hat c).
\end{equation}
\noindent 
Apart from characterizing geometric properties of lattice
quantum states, it is needed in the construction of the
quantum Hamiltonian of the real connection approach.

Loll \cite{Loll1995b} showed that a lattice Wilson loop state has to have
intersections of valence at least 4 in order not to be annihilated
by the volume operator $\sum_n \sqrt{\hat{\det p(n)}}$. 
This result is independent
of the choice of gauge group ($SU(2)$ or $SL(2,\C)$).
The volume operator has discrete eigenvalues, and
part of its non-vanishing spectrum, 
for the simplest case of four-valent 
intersections, was first calculated by Loll \cite{Loll1996a}. 
These spectral calculations were confirmed by
De Pietri and Rovelli \cite{DePietriRovelli1996}, who
derived a formula for 
matrix elements of the volume operator on intersections of
general valence (as did Thiemann \cite{Thiemann1996}).

This still leaves questions about the spectrum itself
unanswered, since the eigenspa\-ces of 
$\hat{\det p(n)}$ grow rapidly,
and diagonalization of the matrix representations becomes a technical 
problem.
Nevertheless, one can achieve a better understanding of 
some general spectral properties of the lattice volume operator,
using symmetry properties. 
It was observed in \cite{Loll1996a} that all non-vanishing eigenvalues of
$\hat{\det p(n)}$ come in pairs of opposite sign. Loll subsequently
proved that this is always the case \cite{Loll1997e}.
A related observation concerns the need for imposing an 
operator condition $\hat{\det p}>0$ on physical states 
in non-perturbative quantum gravity \cite{Loll1997e}, a condition which
distinguishes its state space from that of a gauge theory already
at a kinematical level. 

The symmetry group of the cubic three-dimensional lattice is 
the discrete octagonal group $\cal O$, leaving
the classical local volume function $\det p(n)$ invariant. 
Consequently, one can find a set of operators that commute among 
themselves and with the action
of the volume operator, and simplify its spectral analysis
by decomposing the Hilbert space into the corresponding
irreducible representations \cite{Loll1997f}. 
This method is most powerful when applied to states which are
themselves maximally symmetric under the action of $\cal O$,
in which case it leads to a dramatic reduction of the
dimension of the eigenspaces of $\hat{\det p(n)}$.

In addition to the volume operator, one may define 
geometric lattice operators measuring areas and lengths 
\cite{Loll1997a,Loll1997b}.
They are based on (non-unique) discretizations of the continuum 
spatial integrals 
of the square root of the determinant of the metric induced on 
subspaces of dimension 2 and 1.
For the case of the length operator, operator-ordering
problems arise in the quantization.

\subsection{The real dynamics}
To avoid problems with the non-compactness of
the gauge group and the formulation of
suitable ``quantum reality conditions'', Barbero 
\cite{Barbero1995} advocated to
use a real su(2)-connection formulation for Lorentzian continuum gravity.
This can be achieved, at the price of
having to deal with a more complicated Hamiltonian constraint. 
Loll \cite{Loll1996b,Loll1997c} translated the real
connection formulation to the lattice and studied some of the 
differences that arise in comparison with the complex approach.
Adding for generality a cosmological constant term, this leads to a 
lattice Hamiltonian 
\begin{equation}\label{Z.9}
H^{\rm latt}=
\sum_{n}N^{\rm latt}(n)\; \big[ {\cal H}_{\rm kin}^{\rm latt}+
{\cal H}_{\rm pot}^{\rm latt}+\sqrt{\lambda} G\sqrt{\det p(n)} \big],
\end{equation}
\noindent where schematically ${\cal H}_{\rm kin}^{\rm latt}=
(\det p)^{-1/2}{\rm Tr}\,(U\tau)p^{2}$, ${\cal H}_{\rm pot}^{\rm 
latt}=(({\rm Tr}\; (\tau Up\tau U)-p)^{2}p^{6}$  $(\det p)^{-5/2}$.
This regularized Hamiltonian is well-defined on states with
$\hat{\det p}\not=0$, but its functional form is not simple. 
The negative powers of the determinant of the metric 
can be defined in terms of the spectral resolution of $\hat{\det p}$.
The type of representation and regularization enables one to
handle this non-polynomiality.

\subsection{Summary}
Some progress has been achieved in Hamiltonian lattice gravity, 
using discrete analogues of the Ashtekar variables.
The quantization program is still
open-ended, and no physically non-trivial solutions to the
Wheeler-DeWitt equation are known. The fact that one
can study and evaluate geometric operators provides
useful characterization of quantum states. 

An analysis of the spectrum of the volume operator
is crucial for handling the non-polynomial terms
in the quantum Hamiltonian constraint. 
It will be necessary to
find a suitable truncation or approximation to simplify further
the spectral
analysis of the Wheeler-DeWitt operator.
A suitable quantum analogue of the continuum limit $a\rightarrow 0$
has not yet been established, and in this regard the Hamiltonian
ansatz does not go beyond the results obtained in the
Lagrangian formulations described earlier. 

Why should one bother with a Hamiltonian quantization at all?
Typical quantities one wants to study in a discrete path-integral
approach to gravity are transition amplitudes between three-geometries
on different spatial slices. This is not complete without a
specification of the corresponding quantum states, which are
in principle elements of Hilbert spaces of discrete three-geometries
of the type described above.

\section{Quantum Regge Calculus}

\subsection{Path integral for Regge calculus} 
A path-integral quantization of 4d Regge calculus was first
considered in the early eighties
\cite{RocekWilliams1981,Frohlich1981,Cheegeretal1982}. 
This approach goes back to 
Regge \cite{Regge1961}, who proposed to approximate Einstein's 
continuum theory by a simplicial discretization of 
the metric space-time manifold and the gravitational
action. Its local building blocks are
four-simplices $\sigma$. The 
metric tensor associated with each simplex is expressed as
a function of the squared edge lengths $l^{2}$ of $\sigma$,
which are the dynamical variables of this model. 
For introductory material on classical Regge calculus and
simplicial manifolds, see
\cite{Misneretal1973,Sorkin1975,Frohlich1981,Hartle1985a,Hamber1986,
Williams1992,TuckeyWilliams1992,Ambjornetal1997b}; various
quantum aspects are reviewed in
\cite{Hamber1986,Hamber1990,Hamber1992b,Hamber1992c,Hamber1994a,
Isham1987,Berg1989,Menotti1989,TuckeyWilliams1992,Williams1995}.

One may regard a Regge geometry as a special case of a {\it
continuum} Rieman\-nian manifold, a so-called piecewise 
flat manifold, with a flat metric in the interior of
its 4-simplices $\sigma$, and singular curvature assignments to its
two-simplices $b$ (the bones or hinges).

The Einstein action with cosmological
term in the Regge approach is given by
\begin{equation}\label{X.1}
S^{\rm Regge}[l_{i}^{2}]=\sum_{{\rm bones}\; b} 
V_{b}\   (\lambda - k\ \frac{A_{b}\delta_{b}}{V_{b}})
\sim \int d^{4}x\; \sqrt{g}\ ( \lambda-\frac{k}{2} R),
\end{equation}
\noindent where $k=\frac{1}{8\pi G}$, 
$A_{b}$ is the area of a triangular bone,
$\delta_{b}=2\pi-\sum_{\sigma\supset b}\theta (\sigma,b)$
the deficit angle there, and $V_{b}$ a local
four-volume element. $\theta (\sigma,b)$ is the angle between
the two 3-simplices of $\sigma$ ($\pi$ minus the angle between
their inward normals) intersecting in $b$. Hartle and Sorkin
\cite{HartleSorkin1981} have generalized (\ref{X.1}) to the case 
of manifolds with
boundary. The boundary contribution to the action is
given by
\begin{equation}\label{X.2}
\sum_{b\subset {\rm boundary}}A_{b}\psi_{b} 
\sim \int d^{3}x\; \sqrt{g^{(3)}}\ K, 
\end{equation}
\noindent where $\psi_{b}$ is the angle between the normals
of the two three-simplices meeting at $b$. 
The Euclidean path integral on a finite simplicial complex of 
fixed connectivity takes the form 
\begin{equation}\label{X.3}
Z(k,\lambda)=\int {\cal D}l\;e^{-S^{\rm Regge}[l_{i}^{2}]}
\end{equation}
\noindent with $\int{\cal D}l$ representing the discrete analogue
of the sum over all metrics.
A crucial input in
(\ref{X.3}) is the choice of an appropriate measure ${\cal D}l$. 
In general, a cutoff is required for both short and long edge
lengths to make the functional integral 
convergent \cite{Frohlich1981,Cheegeretal1982,Ambjornetal1997c}. One is
interested in the behaviour of expectation values of local
observables as the simplicial complex becomes large, and the existence
of critical points and long-range correlations, in a
scaling limit and as the cutoffs are removed. 

A first implementation of these ideas was
given by Ro\v cek and Williams \cite{RocekWilliams1981,
RocekWilliams1984}. They obtained a simplicial 
lattice geometry by subdividing
each unit cell of a hypercubic lattice into simplices.
Their main result was to rederive the continuum free propagator 
(see also \cite{Feinbergetal1984} for related results) in the limit
of weak perturbations about flat space.
This calculation can be repeated for Lorentzian
signature \cite{Williams1986}.
 
Some non-perturbative aspects of the path integral were investigated
in \cite{RocekWilliams1984} 
(see also \cite{RocekWilliams1982}). In this work, 
discrete analogues of space-time diffeomorphisms are defined 
as the local link length transformations which leave the
action invariant, and go over to translations in the flat case.
It is argued that an approximate invariance should exist in 4d.
One may define analogues of local conformal transformations on 
a simplicial complex
by multiplication with a positive scale factor at each vertex,
but the global group property is incompatible with the
existence of the generalized triangle inequalities. 

\subsection{Higher-derivative terms} 
Simplicial analogues of higher-derivative terms were introduced in
\cite{HamberWilliams1984,HamberWilliams1986}. 
In the continuum, with an appropriate choice of coupling constants,
their inclusion makes the path integral less ill-behaved.
The simplest
higher-derivative term in Regge calculus is given by
\begin{equation}\label{X.4}
\sum_{ b} \frac{A_{b}^{2}\delta_{b}^{2}}{V_{b}}
\sim \int d^{4}x\ \sqrt{g}\ R_{\mu\nu\rho\sigma}R^{\mu\nu\rho\sigma},
\end{equation}
\noindent with $V_{b}$ denoting the local Voronoi four-volume at $b$.
The fact that (\ref{X.4}) should not be identified with
$\int d^{4}x\ \sqrt{g}\ R^{2}$ is less surprising in the
light of the classical result \cite{Cheegeretal1982,Cheegeretal1984}, 
that Regge's
expression $A_{b}\delta_{b}$ for the
scalar curvature (for $d>2$) converges to
its continuum counterpart not pointwise, but only after integration,
i.e. ``in the sense of measures'' (see also \cite{FriedbergLee1984,
Feinbergetal1984}, where similar convergence properties 
were studied by using an imbedding into a sufficiently
large vector space $\R^{N}$). 

More complicated higher-curvature terms can in
principle be constructed, using a simplicial analogue 
of the Riemann tensor (see, for example, \cite{Regge1961,Hamber1986,
HamberWilliams1986,Brewin1988}), but have up to now not 
been used in numerical simulations. 
A related proposal by Ambj\o rn et al \cite{Ambjornetal1997a} is to
include terms in the action that depend on higher powers of
the deficit angle $\delta_b$, as well as terms containing 
powers of the solid angle $\delta_v$ at a vertex.
The introduction of local vierbeins and parallel transporters
is also necessary if one considers
fermion coupling \cite{Frohlich1981,Ren1988}.

\subsection{First simulations}
The first numerical studies of the Regge action
were undertaken by Berg \cite{Berg1985,Berg1986}
and Hamber and Williams \cite{HamberWilliams1985,HamberWilliams1986}. 
Berg performed a Monte-Carlo simulation of the pure-curvature
action for hypercubic $2^{4}$- and $3^{4}$-lattices with simplicial
subdivision (see \cite{Berg1989}
for a description of the method). 
He used the scale-invariant measure
${\cal D}l=\prod_{i} \frac{dl_{i}}{l_{i}}$. 
To avoid the divergence
that results from rescaling the link lengths, 
he kept the total volume constant, by
performing an overall length rescaling of all links after each move.
This amounts to fixing a typical
length scale $l_{0}:=(v_{0})^{\frac{1}{4}}$, 
where $v_{0}$ is the
expectation value of the 4-simplex volume. 

For $k=0$, he found a
negative average curvature $\langle R\rangle$, and some evidence for a
canonical scaling behaviour of lengths, areas and volumes.
For $k=\pm 0.3$, he
obtained a negative (positive) average deficit angle $\langle\delta
\rangle$
and a positive (negative) $\langle R\rangle$.
A more detailed analysis for $0\leq k\leq 0.1$ 
led Berg \cite{Berg1986} to conclude that there exists a critical
value $k^{c}$ (presumably a first-order
transition \cite{Berg1989}), below which $\langle R\rangle$ is convergent, 
whereas above it diverges.
(Myers \cite{Myers1992} has conjectured that it 
may be possible to perform a similar analysis for Monte-Carlo data 
for the Lorentzian action.)

By contrast, Hamber and Williams \cite{HamberWilliams1985} simulated the 
higher-derivative action 
\begin{equation}\label{X.5}
S[l_{i}^{2}]=\sum_{b}\ [\lambda V_{b}-k\
\delta_{b}A_{b}+a\ A_{b}^{2}\delta_{b}^{2}/V_{b}]
\end{equation} 
\noindent on $2^{4}$- and $4^{4}$-lattices,
using a time-discretized form of the Langevin evolution
equation (see also \cite{Hamber1986}). 
They employed the scale-invariant measure
${\cal D}l=\prod_{i} \frac{dl_{i}}{l_{i}}F_{\epsilon}(l)$,
where $F_{\epsilon}$ enforces an ultra-violet cutoff
$\epsilon$.
They investigated the average
curvature ${\cal R}\sim \langle R\rangle$ and squared curvature 
${\cal R}^{2}$
(scaled by powers of $\langle l^2 \rangle$ to make them dimensionless), as
well as $\langle \delta_{b}^{2}\rangle$ and 
$\langle V_{b}\rangle /\langle l^{2}\rangle$. 
For $\lambda=k=a=0$, one finds a negative $\cal R$ and a
large ${\cal R}^{2}$, indicating a rough geometry. 
For small $a$, one observes a sudden decrease in ${\cal R}^{2}$,
as well as a jump from large positive to small negative values
of $\cal R$ as $\lambda$ is increased.
For large $a$, 
$\cal R$ is small and negative, and the geometry appears to be
smooth. 
Like Berg, they advocated a fundamental-length scenario, where the 
dynamically determined average link length provides an effective
UV-cutoff.

\subsection{The phase structure}
Further evidence for a transition between a region of
rough and smooth geometry comes from Monte Carlo simulations by
Hamber \cite{Hamber1991b,Hamber1992a} 
on $4^{4}$- and $8^{4}$-lattices, this time
with the lattice 
${\cal D}l=\prod_{i} l_{i}dl_{i}$
(see also \cite{Hamber1993b} for a summary of results, and
\cite{Hamber1991a} for more details on the method).
There is a value $k_{c}$ at which the average curvature
vanishes. For $k>k_{c}$,
the curvature becomes large and the simplices
degenerate into configurations with very small volumes. 
He performed a simultaneous fit for $k_{c}$, $A_{\cal R}$
and $\delta$ in the scaling relation 
\begin{equation}\label{X.6}
{\cal R}(k)\buildrel{k\rightarrow
k_{c}}\over\sim A_{\cal R}(k_{c}-k)^{\delta}.
\end{equation}
\noindent This leads to a scaling exponent $\delta\approx 0.60$,
with only a weak dependence on $a$. 
There are even points with $a=0$ that lie in the well-defined, smooth phase.
Hamber also investigated the curvature and volume susceptibilities
$\chi_{\cal R}$ and $\chi_{V}$. At a continuous phase transition, 
$\chi_{\cal R}$ should diverge, reflecting long-range correlations
of a massless graviton excitation. The data obtained are
not incompatible with such a scenario, but the
extrapolation to the transition point $k_{c}$ seems somewhat
ambiguous. On the other hand, one does not expect 
$\chi_{V}$ to diverge at $k_{c}$, which is corroborated
by the simulations. 

\subsection{Influence of the measure}
For the pure Einstein action, (\ref{X.1}) with $\lambda=0$,
some differences between the $M_{1}=dl/l$- 
and the $M_{2}=l dl$-measure 
were investigated on a $4^{4}$-lattice by Beirl et al 
\cite{Beirletal1992a,Beirletal1992b}.
A constant-volume constraint 
was used for simulations with the $M_{1}$-measure, and a
cutoff $l^{2}<const$ for $M_{2}$. 
A study of the $k$-dependence of bulk geometric quantities agreed 
with previous 
simulations, wherever applicable.
For small $k$, both measures lead essentially to identical
results. For the $M_{1}$-measure only, one finds that 
$\langle R\rangle /\langle V\rangle$ and $\langle A\rangle$ 
exhibit a jump at some $k_{0}$, 
due to the formation of spikes (isolated long link lengths, 
with the areas staying small).
From this, and the study of edge length distributions,
one concludes that 
the DeWitt-type measure $M_{2}$ is generally better behaved.

\subsection{Evidence for a second-order transition?}
The same $l dl$-measure was used by Hamber in an extension 
of a previous simulation of the higher-derivative 
action (\ref{X.5}) \cite{Hamber1993a}, for lattice sizes of up to $16^{4}$. 
At the transition point ($k=k_{c}\simeq 0.244$,
$\lambda=1$, $a=0.005$), the distributions of edge lengths, volumes
and curvatures are smooth and Gaussian-like, and the average
curvature vanishes. 
The location of $k_{c}(a)$ from
fits (\ref{X.6}) to the average curvature coincide with those
from \cite{Hamber1992a}, leading to
$\delta\simeq 0.626$. The data at $a=0$ do
not seem to match this interpretation. This leads to the tentative
conclusion that only for sufficiently large $a$ the observed
transition is of second order. It is in general
``difficult to entirely exclude the presence of a weak first-order
transition, if it has a very small latent heat''.
For $a=0.005$, one finds
some evidence for a decrease in the fractal dimension as
$k$ grows.

\subsection{Avoiding collapse}
Starting from the observation that there exist Regge configurations
with $R_{\rm tot}\rightarrow\infty$ and $V_{\rm tot}\to 0$, 
Beirl et al \cite{Beirletal1993,Beirletal1994d} investigated the influence 
of a cutoff $f$ on the fatness of a simplex $\sigma$, defined by 
\begin{equation}\label{X.7}
\phi_{\sigma}:=\frac{V_{\sigma}^{2}}{\max_{l \in \sigma}l^{8}}\geq f >0.
\end{equation}
\noindent 
Such a uniform shrinking of simplices is known to be necessary 
in order for piecewise
flat manifolds to approach their continuum counterparts 
\cite{Cheegeretal1984}.
On the $4^{4}$-lattice, with the $l dl$-measure and $\lambda=0$,
they determined ${\cal R}(k)$ 
for decreasing $f$. For small $k$,
the choice of $f$ seems to have only little influence, but
towards the transition point $k_{c}$, 
$\langle\phi_{\sigma}\rangle$ simultaneously decreases.                                                       
Next they investigated a variety of measures of the form
$l^{2\sigma -1}dl$
(see also \cite{Beirletal1994a}). 
For $\sigma\leq 1$, only a
mild $\sigma$-dependence of $\cal R$ is observed as 
$k\rightarrow k_{c}$. However, for $\sigma=1.5$, there are
significant differences for the entire range of $k$,
and some evidence that the geometry degenerates.

\subsection{Two-point functions}
To understand the nature of the possible
excitations at the phase
transition, one needs to study correlation functions in the
vicinity of $k_{c}$, which is difficult numerically.
Some data are available on the connected
correlation functions of the curvatures and the volumes 
at fixed geodesic distance $d$, $G_{R}(d)$ and $G_{V}(d)$, for
lattice sizes $\leq 16^{4}$
\cite{Hamber1994b}, using a scalar field propagator to
determine $d$. Both correlators were
measured at various $k$-values, leading to similar results
for both $a=0$ and $a=0.005$.
The data, taken for $d\leq 16$ ($\simeq$ 7 lattice
spacings), can be fitted to decaying exponentials. 

Some further data (for $a=0$) were reported
by the Vienna group \cite{Beirletal1994b,Beirletal1997}. These authors used
simply the lattice distance $n$ instead of the true geodesic
distance $d$. 
In \cite{Beirletal1994b}, the measure was taken to be of the form
$l^{2 \sigma-1}dl$.
They looked at $G_{V}(n)$ on $3^{3}\times 8$-
and $4^{3}\times 16$-lattices, for $\lambda=\sigma =1$ and
$\lambda=\sigma =0.1$, and found a fast decay for all investigated
values of $k$, and $n\leq 8$.

\subsection{Non-hypercubic lattices}
Simulations on lattices with irregular link geometry (still with 
$T^{4}$-topology) have been performed by Beirl et al 
\cite{Beirletal1993,Beirletal1994c}. 
They were obtained by adding a few vertices of 
low coordination number to otherwise regular lattices. 
One finds that the average curvature $\langle R\rangle$ increases 
from negative
to positive values, even for $k<k_{c}$, as a result of the formation 
of spikes. 
In \cite{Beirletal1994c}, the averages $\langle l^{2}\rangle$ were 
monitored separately at the regular and the 
inserted vertices. 

The dependence of $\cal R$ on $k$ 
is rather interesting: one observes
{\it two} ``critical'' points, a smaller one $k_{1}$, where the extra 
vertices develop spikes, and a second one $k_{2}$ where the remaining 
vertices follow.
$\cal R$ undergoes a small jump at $k_{1}$, and a larger one at $k_{2}$.
There is also a transition point to a phase with collapsed
simplices at large negative $k$, 
with a jump to large negative $\cal R$. Additional transition
points at negative $k$ were also found in simulations of the
``compactified" Regge action 
$S=\sum_b (-k\ A_b \sin \delta_b\ +
\lambda V_b)$ (this action was discussed in \cite{Frohlich1981};
see also \cite{Caselleetal1989,KawamotoNielsen1991})
and a $\Z_2$-version of Regge gravity \cite{Beirletal1996}.
Correlation functions at those points were
computed in \cite{Beirletal1997} for short distances, 
but no evidence for
long-range correlations was found.

The same authors studied the inclusion of the higher-derivative term 
(\ref{X.4}) in \cite{Beirletal1994c}. On the regular 
lattice, their findings for ${\cal R}(k,a)$ confirmed those by 
Hamber, apart from the fact that they found stable expectation values 
even for positive $\cal R$. 
Inserting irregular vertices pushes $\cal 
R$ to larger values and leads again to the appearance of an 
additional transition point.

\subsection{Coupling to SU(2)-gauge fields}
Berg and collaborators \cite{BergKrishnan1993a,BergKrishnan1993b,
Bergetal1993,Beirletal1995a,Beirletal1995b,Bergetal1996} coupled the 
pure-curvature action geometrically to the Wilson action for 
SU(2)-gauge fields via dimensionless weight factors $W_{b}$,
\begin{equation}\label{X.8}
S=S^{\rm Regge}-\frac{\beta}{2}\sum_{b}W_{b}\ {\rm Re}\ [
{\rm Tr} (1-U_{b})],\qquad W_{b}=const\ \frac{V_{b}}{(A_{b})^{2}},
\end{equation}
\noindent where $U_{b}$ denotes the SU(2)-holonomy around 
$b$ and $\beta$ is proportional to the 
inverse square coupling constant, $\beta=\frac{4}{g^2}$.
One motivation was to understand whether in the well-defined
pure-gravity region, one can 
choose the elementary particle masses to be $\ll m_{\rm Pl}$ as
$\beta\to\infty$, as one might expect for a realistic gravity+matter 
system. This seems a rather distant hope, since in the simulations 
performed so far, the ratio $m_{\rm Pl}/m_{\rm hadron}$ is of order unity. 

Initial computations were performed on a $2\times 4^{3}$-lattice with 
the scale-invariant measure, and at $k=0.01$ \cite{BergKrishnan1993a,
Bergetal1993}, and extended to larger $k$-values 
in \cite{BergKrishnan1993b}. For $k\leq 0.04$, one finds some evidence for
a (first-order?) transition; the region of $\beta$ where the 
transition occurs does not change much with $k$. 
Beirl et al \cite{Beirletal1995a,Beirletal1995b} extended this 
analysis by measuring the static 
potential $V$ of a quark-antiquark pair 
on lattices of size $4\times 6^3$ and $4\times 8^3$. 
With and without gravity, one finds 
both a confined and
a deconfined phase; in the presence of gravity, the
transition occurs at a smaller $\beta$-value. 
More recently, Berg et al \cite{Bergetal1996} have gathered further data on the 
location and stability of the well-defined
phase in the $(k,\beta)$-plane, 
and extracted a string tension for various $\beta$-values. 

\subsection{Coupling to scalar matter}
The coupling of Regge gravity to scalar matter was investigated 
by Hamber and Williams 
\cite{HamberWilliams1994,Hamber1994a}, for $\lambda=1$, $a=0.005$, and
various $k$-values. They considered a single, massive
scalar field, with an action contribution
\begin{equation}\label{X.9}
S_\phi =\frac{1}{2}\sum_{<ij>}V_{ij} \bigg(\frac{\phi_i-\phi_j}
{l_{ij}}\bigg)^2 +\frac{m^2}{2}\sum_i V_i\phi_i^2,
\end{equation}
\noindent where $V_i$, $V_{ij}$ denote the 4-volumes associated
with the vertices and edges. 
Most data were obtained on
the $4^4$-lattice, with the measure $V_l^{(1/30)}\ ldl$.  
The inclusion of the scalar field leads to a slight decrease in
the average volume and edge length, as well as a very slight
decrease in $|{\cal R}|$. They again performed a fit for
${\cal R}$ according to (\ref{X.6}) and found that the 
critical
exponent and the location of $k_c$ were almost unchanged. 

\subsection{Recovering the Newtonian potential}
Hamber and Williams \cite{HamberWilliams1995} used an ansatz 
coming from a weak-field calculation
for the amplitude of two gravitationally interacting
particles separated by a fixed geodesic distance. They used the
$ldl$-measure on a $16^4$-lattice, with $\lambda=1$ and $a=0$.
The potential $V(d)$ is extracted from the connected Wilson-line
to Wilson-line correlator.
From data in $d\in[0,5]$, and for sufficiently large $k$,
one finds an attractive potential $V(d)<0$.
They considered fits to various shapes of the potential $V$, in
order to extract an effective Newton's constant. --
They also suggested an effective action for
$\cal R$ depending on $k,\ \lambda$ and $a$ that shares some 
of the features observed in the simulations.

\subsection{Gauge invariance in Regge calculus?}
A view that has been expressed frequently is 
that away from configurations
with special symmetries,  different edge length assignments
correspond to inequivalent geometries, and in this sense 
Regge calculus possesses no gauge invariance 
(\cite{Sorkin1975,Cheegeretal1982,Hartle1985a}; for a dissenting opinion,
see \cite{RomerZahringer1986}). 

The diffeomorphism
invariance can be recovered in a weak-field perturbation about
flat space, as was shown by Ro\v cek and Williams 
\cite{RocekWilliams1981,RocekWilliams1984},
and there is some evidence for the existence of analogous zero-modes 
in perturbations of regular, non-flat tesselations, at least in
2d \cite{HamberWilliams1997a}.  
Inspired by the perturbative analysis around flat space,
Hamber and Williams \cite{HamberWilliams1997a} argue
that a similar gauge invariance should persist
even if one perturbs around an arbitrary non-flat background. 
They propose as a possible definition for such gauge transformations
local variations $\delta l_i^2$ of the link lengths that leave both  
the local volume and the local curvature terms 
invariant. 

One may hope that in the non-perturbative Regge regime
no gauge-fixing is necessary, since the contributions from
zero-modes cancel out 
in the path-integral representation for operator averages
\cite{Hartle1985a,HamberWilliams1997a}.
Menotti and Peirano \cite{MenottiPeirano1997a,
MenottiPeirano1997b,MenottiPeirano1997c,Menotti1998}, 
following a strategy suggested by Jevicki and Ninomiya 
\cite{JevickiNinomiya1986}, 
have argued vigorously that the functional integral should contain 
a non-trivial Faddeev-Popov determinant.
Their starting point is 
somewhat different from that adopted in the path-integral simulations
(see also \cite{Williams1997}).
They treat piecewise flat spaces as special cases of
differentiable manifolds (with singular metric), with the 
action of the full 
diffeomorphism group still well-defined.
To arrive at a concrete representation for the Faddeev-Popov
term which could be used in simulations seems 
at present out of reach.
 
Recently, Hamber and Williams \cite{HamberWilliams1997a,
HamberWilliams1997b} have argued
that the $ldl$-lattice measure is the essentially unique local
lattice measure over squared edge lengths (this is a special
case of the one-parameter family of local measures of the
form $\prod_\sigma [V(\sigma)]^\nu \prod_{l} ldl$; see also
\cite{Bander1986} for a related derivation). It does of course
require a term with positive cosmological term in the action in
order to suppress long edge lengths.

\subsection{Assorted topics}
Hartle \cite{Hartle1985a,Hartle1986b} 
has suggested to compute the wave functional of
the universe in a simplicial approximation, and 
to evaluate the discrete path integral semiclassically
near stationary points of the Regge action. He investigated
numerically the extrema of the action (\ref{X.1}) on small simplicial
manifolds with topologies $S^{4}$, $\C P^{2}$ and $S^{2}\times
S^{2}$ \cite{Hartle1986a}. The properties of a Hartle-Hawking wave 
functional for a
small complex with an $S^{3}$-boundary 
were studied in \cite{Hartle1989}. 
 
Fr\"ohlich \cite{Frohlich1981} has advocated the need for a proof of
reflection positivity of the Regge path integral,
which one may expect 
to play a role in proving the unitarity of the theory. 
This can be formulated as a condition on the path-integral measure
(including the action) 
under the gluing of two simplicial four-manifolds along a 
three-dimensional boundary.

Other authors have suggested to associate gauge-theoretic 
instead of metric 
variables with the building blocks of a simplicial complex,
for the case of the Poincar\'e group \cite{Caselleetal1989}, the
Lorentz group \cite{KawamotoNielsen1991}, and
for Ashtekar gravity with gauge group $SU(2,\C)$ 
\cite{Immirzi1996,Immirzi1997}, 
and to reformulate the quantum theory in terms of them. 

Hamiltonian 3+1 versions of Regge calculus have been studied 
classically (see \cite{TuckeyWilliams1992} for a review),
but attempts to quantize them have not
progressed very far.
One meets problems with the definition of the
constraints and the (non-)closure of their Poisson algebra.
A recent proposal for constructing a
canonical quantum theory is due to M\"akel\"a \cite{Makela1994}, who
constructed a simplicial
version of the Wheeler-DeWitt equation, based on the use of area 
instead of length variables (which however are known to be
overcomplete). 
In a similar vein, Khatsymovsky \cite{Khatsymovsky1994} has suggested that 
the operators measuring spatial areas ought to have a discrete
spectrum.

\subsection{Summary} 
Quantum Regge calculus is based on the well-explored
classical discretization of the Einstein action due to
Regge. Its weak-field 
limit around flat space agrees with
the continuum result. 
Numerical simulations of the Euclidean path integral 
indicate the existence of
a well-defined phase with small (negative) average curvature
for sufficiently small $k$ and sufficiently large $\lambda$, 
even in the absence of higher-order curvature terms.
Hamber \cite{Hamber1992a,Hamber1993a} has found some evidence 
for a second-order phase
transition in the presence of a small higher-order
derivative coupling, with a vanishing average curvature at the
transition point. These findings have not been confirmed
by other groups. The recent controversy in the
dynamical triangulations approach 
teaches us to treat this issue with some caution. 

Almost all simulations have been done on hypercubic, subdivided
lattices with $T^4$-topology, which may
introduce a systematic bias in the results. 
There is evidence that the choice of measure plays a role in the appearance
and suppression of singular geometries, so-called spikes. 
The study of irregular lattices suggests
a direct link of the transition points with the appearance of such
spikes. 
This feature is reminiscent of the appearance of
singular structures in dynamical triangulations. 
The coupling of a single scalar or SU(2)-gauge field seems to have
little influence on the phase structure of the gravitational sector.

\section{Dynamical triangulations}

\subsection{Introduction}
This quantization approach has received a lot of attention 
since the early nineties \cite{AgishteinMigdal1992a,AgishteinMigdal1992b,
AmbjornJurkiewicz1992}, inspired by
analogous studies in two-dimensional gravity, where 
dynamical triangulation methods
have been a valuable tool in complementing
analytical results (see, for example, \cite{David1995,Ambjorn1996}).
I will here exclusively concentrate on the 4d results. Other
overview material is contained in \cite{Ambjornetal1992,Jurkiewicz1993,
Wheater1994,Ambjorn1995a,Ambjorn1995b, 
Ambjornetal1995,BrugmannMarinari1995b,Catterall1996,Krzywicki1996,
Johnston1997,Ambjornetal1997b,Ambjornetal1997d,Bowick1998}.

Dynamical triangulations are a variant of quantum Regge calculus,
where the dynamical variables are not the edge lengths of a
given simplicial complex, but its connectivity. A precursor is
Weingarten's \cite{Weingarten1982} 
prescription for computing transition amplitudes
between three-geometries, by summing over all
interpolating four-geometries, built from equilateral 4d hypercubes
living on an imbedding $p$-dimensio\-nal hypercubic lattice with
lattice spacing $a$. Evaluating the Einstein action on such a
configuration amounts to a simple counting of hypercubes of
dimension 2 and 4, c.f. (\ref{Y.1}). 

To avoid a potential overcounting
in the usual Regge calculus, R\"omer and Z\"ahrin\-ger 
\cite{RomerZahringer1986}
proposed a gauge-fixing procedure for
Regge geometries. 
They argued for an essentially unique association of Riemannian
manifolds and equilateral triangulations that in a certain
sense are best approximations to the continuum manifolds. 
The resulting ``rigid Regge
calculus'' is essentially the same structure that nowadays
goes by the name of ``dynamical triangulations''.
In this ansatz one studies the statistical
mechanical ensemble of triangulated 
four-manifolds with fixed edge lengths, weighted by 
the Euclideanized Regge action, with a cosmological 
constant term, and optionally higher-derivative contributions.
Each configuration represents a discrete geometry,
i.e. the discrete version of a Riemannian four-metric modulo 
diffeomorphisms. 
At least for fixed total volume, the state sum converges for
appropriate values of the bare coupling constants, if one restricts
the topology (usually to that of a sphere $S^{4}$). 

\subsection{Path integral for dynamical triangulations}
Denoting by $\cal T$ the set of all triangulations of the four-sphere,
the partition function for the model is given by
\begin{equation}\label{Y.1}
Z(\kappa_{2},\kappa_{4})=\sum_{T\in{\cal T}} 
\frac{1}{C(T)} e^{-S[T]},\qquad
S[T]=-\kappa_{2}N_{2}(T)+\kappa_{4}N_{4}(T),
\end{equation}
\noindent where $N_{2}$ and $N_{4}$ denote the numbers of 2- and
4-simplices contained in the simplicial manifold $T$, and $C(T)$ 
is the order of the automorphism group of $T$. 
One may think of (\ref{Y.1}) as a grand canonical ensemble, with chemical
potential $\kappa_{4}$. It is related to the canonical ensemble
with fixed volume, $Z(\kappa_{2},N_{4})$, by a Legendre
transform
\begin{equation}\label{Y.2}
Z(\kappa_{2},\kappa_{4})=\sum_{N_{4}}e^{-\kappa_{4}N_{4}}Z(\kappa_{2},N_{4})
\equiv \sum_{N_{4}}e^{-\kappa_{4}N_{4}}\sum_{T\in {\cal T}(N_{4})}
e^{\kappa_{2}N_{2}(T)}.
\end{equation}
\noindent The metric information is encoded in the connectivity of the
simplicial decomposition, since the individual 4-simplices are
assumed equilateral, with the edge length $a$ set to 1.

To understand the simple form of 
the action $S$, recall that the curvature term
in Regge calculus (c.f.(\ref{X.1}) is represented by 
$ \sum_{b}2 \delta_{b}A_{b}$, which
for fixed edge length is proportional to 
$(c_4 N_{2}-10 N_{4})$. The constant 
$c_4=  2\pi/\arccos\frac{1}{4} =4.767$ is determined
from the condition that a triangulation of flat space
should have average vanishing curvature \cite{AgishteinMigdal1992a,
AmbjornJurkiewicz1992}. (Because the four-simplices $\sigma$ 
are equilateral, zero
curvature can only be achieved upon averaging. This
explains the absence of a conventional
perturbation theory around flat space.)   
The cosmological term is represented by 
$\lambda N_{4} V(\sigma)\sim \lambda N_{4}$.
It is sometimes convenient to re-express $N_{2}$ as a function
of $N_{0}$, using $N_{2}=2N_{0}+2N_{4}-4$,
valid for the $S^{4}$-topology. The 
corresponding partition function is $Z(\kappa_{0},\kappa_{4})$
(where $\kappa_{0}=2\kappa_{2}$).  

\subsection{Existence of an exponential bound?}
As a consequence of identities and inequalities satisfied by
the $N_{i}$ (\cite{Ambjornetal1997b} contains a detailed
discussion), 
the action (\ref{Y.1}) is bounded above and below for fixed volume $N_{4}$. 
If the number of configurations for fixed $N_{4}$, is exponentially
bounded as $N_{4}\rightarrow\infty$, that is, $Z(\kappa_{2},N_{4})$
grows at most as $Z(\kappa_{2},N_{4})\sim e^{const\,N_{4}}$, 
there is a ``critical line'' $\kappa_{4}=\kappa_{4}^{c}(\kappa_{2})$
in the $(\kappa_{2},\kappa_{4})$-plane, where for fixed $\kappa_{2}$,
$Z(\kappa_{2},\kappa_{4})$ converges for 
$\kappa_{4}>\kappa_{4}^{c}(\kappa_{2})$. True critical behaviour
may be found by approaching suitable points on this line from the
region {\it above} the line, where $Z$ is well defined. 

Doubts on the existence of an exponential bound were raised
by Catterall et al \cite{Catteralletal1994b}, 
who considered the behaviour of 
$\Omega$ in 
$Z(\kappa_{0},\kappa_{4})=\sum_{N_{4}}e^{-\kappa_{4}N_{4}} 
\Omega (\kappa_{0},N_{4})$.
Their data
(taken for $N_{4}\leq 32k$) were consistent with a leading factorial
behaviour $\Omega\sim (N_{4}!)^{\delta}$.
The same scenario was favoured by de Bakker
and Smit \cite{deBakkerSmit1994b}, who performed further investigations of
$\kappa_{4}^{c}$. 
Subsequently, Ambj\o rn and Jurkiewicz \cite{AmbjornJurkiewicz1994} and
Br\"ugmann and Marinari \cite{BrugmannMarinari1995a} added
further data points at $N_4=64k$ and $N_{4}=128k$ respectively.
Their numerical results, as well as those by 
Catterall et al \cite{Catteralletal1996a}, who employed an 
alternative method for measuring $\Omega$, 
favour the existence
of an exponential bound, although they cannot claim to be conclusive.

There have also been theoretical arguments for the existence of 
an exponential bound, based on the proofs of such bounds for the 
counting of minimal geodesic ball coverings of Riemannian spaces of
bounded geometry \cite{CarforaMarzuoli1995a,Bartoccietal1996}, and the
counting of discrete curvature assignments to unordered sets of bones
\cite{Ambjornetal1997b}.

\subsection{Performing the state sum}
The partition function is evaluated numerically with the help
of a Monte-Carlo algorithm
(see \cite{Brugmann1993b,Bilkeetal1995,Catterall1995}
for details). There is
a set of five topo\-lo\-gy-preserving moves which change a 
triangulation locally, and which are ergodic in the 
grand canonical ensemble 
\cite{Pachner1986,Pachner1991,GrossVarsted1992} (see
also \cite{Ambjornetal1997b} for a discussion).
No ergodic finite set of topology- {\it and} volume-preserving moves
exists for generic four-dimensional manifolds.
This prevents one from using the canonical (i.e. volume-preserving)
ensemble. If $S^{4}$ is not
algorithmically recognizable in the class of all piecewise
linear (or smooth) 4d manifolds,  
the numerical
simulations may miss out a substantial part of the state space
because of the absence of ``computational ergodicity''
\cite{NabutovskyBen-Av1993}.

An attempt was made by Ambj\o rn and Jurkiewicz 
\cite{AmbjornJurkiewicz1995a} to link the non-recognizability 
to the presence of large-$N_{4}$ barriers, which should manifest 
themselves as an obstacle to cooling down an large initial random
triangulation to the minimal $S^{4}$-configuration. No such barriers were 
found for system
sizes $\leq 64k$, but unfortunately they were also absent for an
analogous simulation (for $N_{5}\leq 32k$) performed by de Bakker 
\cite{deBakker1995b} for $S^{5}$, which is not recognizable. 

Since the local moves alter the volume, one works in practice with
a ``quasi-canonical" ensemble, i.e. one uses  
the grand canonical ensemble
$Z(\kappa_{2},\kappa_{4})$, but adds a potential term to the
action so that the only
relevant contributions come from states in an interval 
$[V-\Delta N_{4},V+\Delta N_{4}]$ around the target volume $V$. 
There have been several cross-checks which have found no 
dependence of the results on the width and shape of the potential
term \cite{Catteralletal1996b,Bilkeetal1997a}, but the lattice
sizes and fluctuations may still be too small to detect a
potential failure of ergodicity, c.f. \cite{Bilkeetal1997b}. 

To improve the efficiency of the algorithm, Ambj\o rn and Jurkiewicz
\cite{AmbjornJurkiewicz1995b} used additional
global (topo\-lo\-gy-preserving) ``baby universe surgery'' moves, by 
cutting and gluing pieces of the simplicial complex .
In the branched polymer
phase, one can estimate the entropy exponent $\gamma$, assuming a
behaviour of the form $Z(\kappa_2,N_4)=N_4^{\gamma(\kappa_2)-3}
e^{\kappa_4^c(\kappa_2)N_4}\times (1+O(1/N_4))$, by counting
baby universes of various sizes. At the transition point, one
finds $\gamma(\kappa_2^c)\approx 0$ \cite{Ambjornetal1993b,
AmbjornJurkiewicz1995b,Egawaetal1998a}. More recently,
Egawa et al \cite{Egawaetal1998b} have reported a value of $\gamma\approx
0.26$.

\subsection{The phase structure}
Already the first simulations by Agishtein and Migdal
\cite{AgishteinMigdal1992a,AgishteinMigdal1992b} and 
Ambj\o rn and Jurkiewicz \cite{AmbjornJurkiewicz1992}
of dynamically triangulated 4d gravity
exhibited a clear two-phase structure. After tuning to the
infinite-volume or critical line $\kappa_4^c(\kappa_2)$,
one identifies two regions, $\kappa_{2}<\kappa_{2}^{c}$ and
$\kappa_{2}>\kappa_{2}^{c}$. 
The critical value $\kappa_{2}^{c}$ 
depends on the volume $N_{4}$, and it was conjectured
that it may move out to $+\infty$ as $N_{4}\rightarrow\infty$
\cite{Catteralletal1994a,deBakkerSmit1994b}, but it was later shown 
to converge to a finite value \cite{AmbjornJurkiewicz1995b}. 
One can characterize the region with small
$\kappa_{2}<\kappa_{2}^{c}$ 
as the hot, crumpled, or condensed phase. It has small negative
or positive curvature, large (possibly infinite) Hausdorff
dimension $d_{H}$ and a high connectivity.
By contrast, for $\kappa_{2}>\kappa_{2}^{c}$ one is in the
cold, extended, elongated, or fluid phase. It has large positive curvature,
with an effective tree-like branched-polymer geometry,
and $d_{H}\approx 2$.

The location of the critical point on the infinite-volume line 
may be estimated from the peak in the 
curvature susceptibility
$\chi(\kappa_{2},N_{4})=(\langle N_{2}^{2}\rangle -
\langle N_{2}\rangle^{2})/N_4$
\cite{AgishteinMigdal1992a,AgishteinMigdal1992b,AmbjornJurkiewicz1992, 
Varsted1994,Hottaetal1995,deBakkerSmit1995a}, or the node susceptibility
$\chi(\kappa_{0},N_{4})=(\langle N_{0}^{2}\rangle -
\langle N_{0}\rangle^{2})/N_4$
\cite{Catteralletal1994a,Catteralletal1994c}, 
as well as higher cumulants of $N_0$
\cite{Bialasetal1996}. 
In \cite{AmbjornJurkiewicz1995b}, it was suggested that
one may alternatively estimate $\kappa_2^c(N_4)$ by looking
at the behaviour of the entropy exponent $\gamma$,
approaching $\kappa_2^c$ from
the elongated phase. More recently, Catterall et al 
\cite{Catteralletal1998}
have used the fluctuations 
$\chi_0=(\langle\omega_0^2\rangle -\langle\omega_0\rangle^2)/N_4$ 
of the local volume
$\omega_0$ around singular vertices as an order parameter.

\subsection{Evidence for a second-order transition?}
It sometimes seems to be assumed that if one were to
find a continuum theory at a second-order phase transition, it would have 
flat Minkowski space as its ground
state (in spite of the $S^4$-topology), and gravitonic
spin-2 excitations. 
An alternative scenario with a constant-curvature sphere-metric 
has been put forward by de Bakker and Smit 
\cite{deBakkerSmit1995a,deBakkerSmit1996}. 

Agishtein and Migdal \cite{AgishteinMigdal1992a} 
initially reported a hysteresis
in the average curvature $\langle R\rangle(\kappa_2)$, 
indicating a first-order transition. 
However, subsequent authors
\cite{AmbjornJurkiewicz1992,Varsted1992,Varsted1994,Catteralletal1994a,
Catteralletal1994c}
found numerical data not incompatible with the existence
of a second-order transition, and also Agishtein and Migdal
\cite{AgishteinMigdal1992b} retracted their original claim as a result of a
closer examination of the fixed point region. 

One tries to discriminate between a first- and second-
(or higher-)order transition by looking at
the Binder parameter \cite{Varsted1992,Varsted1994,Ambjornetal1993c},
or scaling exponents $\alpha$ governing the scaling behaviour 
$\sim |\kappa_2^c-\kappa_2|^\alpha$ of suitable observables  
\cite{AgishteinMigdal1992b,Varsted1994},
or the peak height of 
susceptibilities as a function of the volume $N_{4}$
\cite{Ambjornetal1993a,Catteralletal1994a,Catteralletal1994c,
AmbjornJurkiewicz1995b,
Bialasetal1996,deBakker1996}. Other scaling relations are
discussed in \cite{deBakkerSmit1995a,deBakkerSmit1995b,Egawaetal1997a,
Egawaetal1997b}.
However, since critical parameters are hard to measure,
and it is difficult to estimate finite-size effects, 
none of the data can claim to be conclusive. 

Some doubts were thrown on the conjectured
continuous nature of the phase transition by Bialas et al 
\cite{Bialasetal1996}, 
who found an unexpected two-peak structure in 
the distribution of
nodes near the fixed point.
This was strengthened further 
by data taken at 64k by de Bakker \cite{deBakker1996}, with an even 
more pronounced double peak
(see also \cite{Bilkeetal1997a}). 
Most likely previous simulations were
simply too small to detect the true nature of the phase transition.
Both Bialas et al and de Bakker observed that
the finite size scaling exponents extracted from the node
susceptibility $\chi_{0}$ grow with volume and may well reach
the value 1 expected for a first-order transition as $V\rightarrow
\infty$.

The origin of this behaviour was further elucidated by Bialas
et al \cite{Bialasetal1997,Bialasetal1998} and Bialas and Burda 
\cite{BialasBurda1998}, who found a simple 
mean-field model that reproduces qualitatively the phase
structure of 4d dynamically triangulated quantum gravity. 
With an appropriate choice of local weights, this model has
a condensed and a fluid phase, with a first-order transition
in between. A similar behaviour was found by Catterall et al
\cite{Catteralletal1998}, who made a related mean-field ansatz,
with the local weights depending on the local entropies
around the vertices. 

\subsection{Influence of the measure}
The inclusion of a local measure term corresponding to 
$(\det g)^\frac{\mu}{2}$ was studied by Br\"ugmann and Marinari 
\cite{Brugmann1993a,Brugmann1993b,BrugmannMarinari1993}. 
This amounts to adding to the action a term of the form
$-\mu\sum_v \ln o(v)/5$, where $o(v)$ is the number of 4-simplices
containing a vertex $v$. \
For $\mu=-5,-1,1,5$, this seems to lead merely to a shift of the
critical line. 
This setting has been revived recently by Renken
\cite{Renken1997a,Renken1997b} 
in the context of a renormalization group analysis.

\subsection{Higher-derivative terms}
Higher-order derivative terms were considered by Ambj\o rn et 
al \cite{Ambjornetal1993a} (see also \cite{Kristjansen1993}), 
who added a term of the form 
\begin{equation}\label{Y.3}
h\, c_4^{-2}\sum_{b} o(b)\bigg(\frac{c_4-o(b)}{o(b)}
\bigg)^2
\end{equation}
\noindent to the action, where again $c_4=4.767$ and
$o(b)$ denotes the number of 4-simplices sharing a bone $b$. 
For $h\in [0,20]$, and with volumes up
to $32k$, no major qualitative changes of the geometrical
observables were found. The inclusion
of the higher-derivative term also does not improve
the behaviour of the average curvature $\langle R\rangle$, which
continues to be positive at the critical point,
whereas from a na\"\i ve comparison with the continuum
theory one would expect it to scale to zero. 
(This is also incompatible with the prediction of Antoniadis et
al \cite{Antoniadisetal1994}, should dynamical triangulations possess 
an infrared stable fixed point.)
De Bakker and Smit \cite{deBakkerSmit1994b} have argued that this may not 
be a reason of concern, since one expects the volume and
curvature terms to mix under renormalization.  

\subsection{Coupling to matter fields}
Ambj\o rn et al \cite{Ambjornetal1993c} considered the influence of both
Ising spins and Gaussian scalar fields on bulk geometric
quantities.
The phase structure remained essentially unchanged, and no
improvement in the scaling behaviour of $\langle R\rangle$
was found. 
Taken together with their results on higher-derivative gravity
\cite{Ambjornetal1993a}, one finds a universal linear dependence
of the cosmological constant $\kappa_4^c(\kappa_2)$, with 
slope $\sim 2.5$.

Coupling to $\Z_{2}$-spin variables $s(l)$ located on edges $l$ was
considered by Ambj\o rn et al \cite{Ambjornetal1994}, who added a Wilson loop
term $S_{\rm W}[s;T]=-\beta \sum_{b\in T}o(b) 
\prod_{l\in b}s(l)$ to the action.
The matter sector behaves largely as expected when $\kappa_{2}$
is varied between the crumpled and the elongated gravity phase.
However, in the common critical region of both sectors,
where {\it a priori} one
might have expected interesting effects, the critical behaviour seems 
to agree with that of the pure gravity system.

More recently, Bilke et al \cite{Bilkeetal1997c} have reported a non-trivial
back-reaction of matter on geometry, when considering 
coupling to several non-compact $U(1)$-gauge fields. 
Their study was in part motivated by a continuum analysis
of the dynamics of the conformal factor of Antoniadis et al
\cite{Antoniadisetal1997}.
Including three fields $U(1)$-fields seems to lead to a total
suppression of the branched polymer phase, which is replaced
by a new weak-coupling phase with negative susceptibility
exponent $\gamma$ and a fractal dimension $\approx 4$.
These are clearly interesting results, but should be treated
with some caution because of the small lattice sizes involved
($N_4\leq 16k$).

\subsection{Non-spherical lattices}
Alternative topologies for the underlying simplicial complex
were considered by Bilke et al \cite{Bilkeetal1997a,Bilkeetal1997b}.
They showed that for topologies $S^{1}\times S^{3}$ and $T^{4}$,
the free energy agrees to leading order with that of $S^{4}$.
Neither the critical value $\kappa_{4}^{c}(\kappa_{2},N_{4})$,
nor the leading contributions to the action density
$\langle N_{2}\rangle/N_{4}$ are changed. 
However, the simulations become
more involved, since the size of the minimal configurations
is increased.

\subsection{Singular configurations}
The singular nature of the geometry in the phase 
below the critical value
$\kappa_{2}^{c}$ can be quantified by the distribution
$\rho(n)$ of the vertex order $o(v)$,
\begin{equation}\label{Y.4}
\rho (n)=\frac{1}{N_{0}} \bigg< \sum_{v}\delta_{o(v),n}\bigg>,
\end{equation}
\noindent first considered by Hotta et al 
\cite{Hottaetal1995}. It
has a continuum part and a separate peak at high vertex order.
One can suppress this effect by adding a term $\sim \sum_{v}
(o(v)-5)^{2}$ to the action
(see also \cite{Ambjornetal1997a} for a discussion of terms of
a similar nature), but this leads to a
simultaneous disappearance of the phase transition.
Hotta et al \cite{Hottaetal1996} have checked that 
for a variety of initial configurations the 
singular structure is a generic feature of the model.

Catterall et al \cite{Catteralletal1996b,Catteralletal1997} 
observed that the pair of
singular vertices form the end points of a singular link.
They also offered
a possible explanation for the formation of these singular
structures: simplices of sufficiently low dimension can
maximize their local entropy by acquiring large local volumes
(see also \cite{Bialasetal1997} for a mean-field argument). 
Catterall et al 
\cite{Catteralletal1998} found {\it two} pseudo-critical points,
$\kappa_{0}^{(1)}$ and $\kappa_{0}^{(2)}$, associated with the
creation of singular vertices and links, which seem to merge
into a single critical point $\kappa_{0}^{c}$ as $V\rightarrow
\infty$. One concludes that
the observed phase transition in the 4d dynamical 
triangulations model is driven by the appearance and disappearance 
of singular geometries.

\subsection{Renormalization group}
There have been attempts to apply renormalization group
techniques, assumpting that the transition is
indeed continuous. 
Burda et al \cite{Burdaetal1995} and
later Bialas et al \cite{Bialasetal1996}
considered the ``cutting of the last generation of minimal-neck baby
universes". This step can only be performed once,
which severely limits the power of the procedure. 
Renken \cite{Renken1997a,Renken1997b} has applied a different blocking move, 
involving node deletion, and 
studied the RG flow using the volume and the vertex order
as observables.

\subsection{Exploring geometric properties}
Since the effective geometry in both phases is rather singular,
different ways of measuring length may lead
to inequivalent definitions of ``dimension". 
A common (local) notion is derived from the
volume of a geodesic ball with radius $r$. Usually the radius
is measured in terms of the geodesic distance $d_{1}$ (the minimal number
of links), or dual geodesic distance $d_{4}$
(the minimal number of links of the dual graph).
Alternatively, one may consider the number $n(r)$ of 4-simplices
in spherical shells of thickness 1 at distance $r$, and define
a fractal dimension by $\langle n(r)\rangle\sim r^{d_{\rm F}-1}$ 
\cite{deBakkerSmit1995a,AmbjornJurkiewicz1995b}.

To extract a global length scale, one may use the averages
$\langle d_{1}\rangle$, $\langle d_{4}\rangle$, or consider the average
``radius of the universe'' $\langle r(T)\rangle$ 
\cite{AmbjornJurkiewicz1992}
to obtain a cosmological Hausdorff dimension $d_{\rm CH}$, or the
``average intrinsic linear extent'' 
$L=\langle V^{-2}\sum_{i,j}d_{4}(i,j,T)\rangle$ 
(see, for example, \cite{Catteralletal1994a,Catteralletal1994c}).
Away from the phase transition, the fractal dimensions associated
with these geometric construction are more or less equivalent
and give $d_{\rm F}=\infty$ in the crumpled phase and $d_{\rm F}\approx 2$
in the elongated phase. It is difficult to measure the dimension
close to the transition point.

It will not be straightforward to interpret the
behaviour of observables (defined in analogy with the continuum
theory), since in most of the phase space the geometry of the
simplicial complex is far from approximating a 
metric 4-manifold. 
In search of a semiclassical interpretation for geometric
observables, an alternative notion of local
curvature for a simplicial manifold was suggested by
de Bakker and Smit \cite{deBakkerSmit1995a}, 
based on a continuum expansion of the 
volume of a geodesic ball.
Assuming furthermore that 
independent of $n$, $n$-volumes of balls with radius $r$ 
behave like regions
on $S^{n}\subset \R^{n+1}$,
they extracted scaling relations for various geometric quantities
for an intermediate range for $r$. This line of thought was
pursued further in \cite{Smit1997}.

Close to the phase transition, one may investigate the behaviour
of test particles (ignoring back-reactions on the geometry).
Comparing the mass extracted from the one-particle propagator
with the energy of the combined system obtained from the
two-particle propagator \cite{deBakkerSmit1994a,deBakkerSmit1996,
deBakkerSmit1997}, one
does indeed find evidence for gravitational binding.

\subsection{Two-point functions}
It is possible to define two-point correlation functions
on random geometries \cite{Ambjornetal1993a,AmbjornJurkiewicz1995b}, 
which are to be thought of as the 
discrete analogues of formal continuum correlators 
\begin{equation}\label{Y.5}
G_{\cal O}(r)=\int_{\frac{{\rm Riem} S^{4}}{{\rm Diff} S^{4}}}
Dg_{\mu\nu} e^{-S}\int d^{4}x\int d^{4}x'\sqrt{\det g(x)}
\sqrt{\det  g(x')}{\cal O}(x){\cal O}(x')\delta (d_{g}(x,x')-r),
\end{equation}
\noindent for local observables ${\cal O}(x)$, with 
$d_{g}$ denoting the geodesic distance with respect
to the metric $g_{\mu\nu}$. 
There is an ambiguity in defining the 
{\it connected} part of the correlator (\ref{Y.5}),
as was pointed out by de Bakker and Smit 
\cite{deBakkerSmit1995b,deBakkerSmit1996}.
Contrary to expectations,
after subtraction of the square of the curvature expectation value, 
the resulting
quantity $\langle RR\rangle (r)-\langle R\rangle^{2}$ 
does not scale to zero with large distances. 
They therefore proposed an alternative definition
of the connected two-point function, by subtracting the square of a
``curvature-to-nothing'' correlator $\langle R\rangle(r)$. 
This definition was compared
in more detail by Bialas \cite{Bialas1997} with a more conventional
notion, as, for example, the one used in \cite{Bialasetal1996}. For the
case of curvature correlators, their behaviour differs significantly,
especially at short distances.

\subsection{Summary}
In the dynamical triangulations approach, one studies the properties of 
a statistical ensemble of simplicial four-geometries \`a la Regge of fixed
edge lengths. By summing over such discrete configurations 
according to (\ref{Y.1}), 
one has implicitly assumed that this leads to a uniform sampling of
the space of smooth Riemannian manifolds. 
There is no obvious weak-field limit, but this is no obstacle in principle
to the path-integral construction. Numerical simulations 
indicate the existence
of a well-defined phase for sufficiently small $\kappa_2$ (inverse
Newton's constant) and a sufficiently large cosmological constant.
For small $\kappa_2$, one finds a ``crumpled" phase, with small
average curvature and a large Hausdorff dimension, and for large
$\kappa_2$ an elongated, effectively two-dimensional polymer phase.
At present, the consensus seems to be that the corresponding
phase transition is of first order, with a finite average curvature
at the transition point. 

Almost all simulations have been done on simplicial manifolds with
$S^4$-topology. Neither the inclusion of factors of $(\det g)^\rho$
in the measure nor the addition of higher-order curvature terms
to the action seem to have a substantial influence on the phase
structure. Also matter coupling to spinorial and
scalar fields does not seem to lead to a change of universality
class, although the inclusion of several gauge fields may have
a more drastic effect. The study of singular structures (vertices
of high coordination number) has led to a qualitative
understanding of the phase structure of the model.

\section{Conclusions and Outlook}

In this review, I have collected a variety of results on discrete
four-dimensional models of quantum gravity, mainly coming from
Euclidean path-integral approaches. Numerical simulations 
have yielded information on the phase
structure of these models, the behaviour of two-point functions and a number
of other properties of their partition functions.
All of the path-integral models 
have some qualitative features in common. They
need a (sufficiently large, positive) cosmological constant
$\lambda$ to be well defined. For
sufficiently small values of Newton's constant $G$, 
one finds a phase
of collapsed geometry, with effective dimension $<4$. In the 
gauge-theoretic model, the metric is degenerate, in Regge calculus,
one finds spiky configurations, and in dynamical triangulations,
the ensemble behaves like that of a branched polymer. 

In all cases, one observes a transition 
on the boundary of this phase, but so far
no convincing evidence of long-range correlations has been found
in its vicinity. 
Within the accuracy of the numerical simulations, this main
conclusion is not altered by 
the inclusion of determinantal factors $(\det g)^{\rho}$
in the measure,
the inclusion of higher-order derivative terms, 
or the addition of matter fields.
{\it Why} does this
happen? Each of the approaches can claim that its state space
represents, at least roughly, an approximation to the space of 
smooth Riemannian metrics or geometries. This leaves only the
path-integral measure as a possible culprit. The measures used
up to now were the simplest ones compatible with considerations
of locality and gauge-invariance.
It seems premature to blame the absence of diffeomorphism invariance
(whose status in the gauge-theoretic formulation
and the Regge calculus program remains unclear), 
since the explicitly 
diffeomorphism-invariant dynamical triangulations approach 
suffers from similar problems. 
Further analytical
insights are needed to understand which modifications of the 
measure would make these models more interacting.

There are a number of loop holes which could change
the picture just presented. It is possible that adding enough
matter of the correct type could have a non-trivial effect, or that
Regge calculus with the inclusion of higher-order curvature terms 
does indeed possess a second-order phase transition.
Since we have very little experience with universality
properties of 4d generally covariant theories, it is not {\it a
priori} clear whether the choice of measure and the initial
restrictions on the lattice geometry can affect the final results. 

One may of course take the attitude that something is fundamentally
wrong with trying to construct a theory of quantum gravity 
via a statistical field theory approach, and that a different
starting point is needed, an obvious candidate being a 
non-perturbative theory of superstrings, or of more general
extended objects. In any case, these different
approaches need not be mutally exclusive, 
and one may therefore take the results of the discrete approaches
presented here as an indication that other attempts of 
constructing quantum gravity non-perturbatively may run into
similar difficulties. 

A further unresolved problem is the ``analytic continuation" of
the path-integral results to Lorentzian signature.
The Hamiltonian ansatz circumvents this problem,
and some progress has been made in the Hamiltonian gauge-theoretic 
discrete approach. Although the kinematical structure is in place
and some information on the constraint algebra has been obtained,
the physical state space has not yet been identified. Its results 
are therefore not sufficiently complete to admit comparison
with the path-integral simulations. 
For the simplicial formulations,
only little is known about their canonical counterparts. 
One would hope that future research will throw further light
on these issues.
  
\vspace{1cm}
\noindent {\bf Acknowledgement.} I am indebted to 
R. Williams, P. Menotti, H. Hamber, W. Beirl and J. Ambj\o rn
for comments and criticism, to E. Schlenk for help with
the references, and to R. Helling for assistance in reformatting
them.  

\newpage

\bibliography{myliv}
\end{document}